\documentclass[11pt]{article}

\usepackage[latin1]{inputenc}
\usepackage{graphicx}
\usepackage{mathrsfs}
\usepackage{amsmath}
\usepackage{amsthm}
\usepackage{amsfonts}
\usepackage{amssymb}
\usepackage{subfig}
\usepackage{color}

\newtheorem{prop}{Proposition}[section]
\theoremstyle{definition}

\title{Bell inequalities as constraints on unmeasurable correlations} 
     
\author{C. Budroni \\ 
{Departamento de F\'{\i}sica Aplicada II, 
Universidad de Sevilla, Sevilla, Spain} \\ \\
{G. Morchio} \\  {Dipartimento di Fisica 
dell'Universit\`a and INFN, Pisa, Italy}}

\begin{document}

\maketitle

\begin{abstract}
The interpretation of the violation of Bell-Clauser-Horne 
inequalities is revisited, in relation with the notion of 
\textit{extension} of QM predictions to 
\textit{unmeasurable correlations}.
Such extensions are compatible with QM predictions 
in many cases, in particular for observables with compatibility relations 
described by tree graphs. This implies classical 
representability of any set of correlations   
$ \langle A_i \rangle$, $ \langle B \rangle$, 
$ \langle A_i B \rangle$, 
and the equivalence of the Bell-Clauser-Horne inequalities 
to a non void intersection between the ranges of values 
for the \textit{unmeasurable correlation} 
$\langle A_1 A_2\rangle$ associated to different choices 
for $B$. The same analysis applies to the Hardy model and to the 
\lq\lq perfect correlations\rq\rq\ discussed by Greenberger, Horne, 
Shimony and Zeilinger. In all the cases, the dependence of an 
unmeasurable correlation on a set of variables allowing for a 
classical representation is the only basis for arguments about 
violations of locality and causality.

\end{abstract}

\newpage
\section*{Introduction and results}
The implications of the violation of Bell inequalities \cite{Bell64} 
in Quantum Mechanics (QM) are still controversial.
On one side, their logical and probabilistic nature has been 
recognized and discussed \cite{Fine}, \cite{GM}, 
\cite{Pit89} and
their fundamental meaning traced back \cite{Pit94} 
to Boole's general notion 
of \textit{conditions of possible experience} \cite{Boole}.
In this perspective, their violation, which excludes representability 
of quantum mechanical predictions by a classical probability 
theory, shows  general and fundamental differences between 
the classical and the quantum notion of event.

On the other, the violation of Bell inequalities is interpreted  
by many authors in terms of \lq\lq non local properties\rq\rq\ of QM; 
an extensive review of such positions is contained in Ref. \cite{Griff} 
and a discussion of their basic logical steps is in
Refs. \cite{Hardy}, \cite{Stapp1}, \cite{Mermin}, \cite{Stapp2}.

In our opinion, in general, the use of Bell inequalities for the 
analysis of QM and of its interpretation is in a sense incomplete 
because their violations simply amounts to a negative result, 
i.e. the inconsistency between QM and \textit{the whole set} 
of assumptions entering in their derivation. 

The present paper is an attempt to reconsider the situation from a more 
constructive point of view, beginning with QM predictions, trying to 
represent them by classical probability models in a sequence of steps 
and asking which step may fail.  

We recall that adopting a classical probability model 
exactly amounts to assuming that \textit{all variables together 
take definite, even if possibly unmeasurable, values}, with
definite probabilities.
 
Since a classical probability model is equivalent to its set of 
predictions for \textit{all} correlations, while QM only predicts
correlations between \textit{compatible} observables, 
classical representability also amounts to an extension of QM
predictions to \textit{unmeasurable correlations}, e.g., 
correlations between two components of the spin 
\textit{of the same particle}, 
or polarization of \textit{the same photon} along different directions.

A central role is therefore played by the notion of 
\textit{extension  of QM predictions to unmeasurable correlations}. 
We shall use basic and general constraints 
on such \textit{extensions} to discuss the 
violation of Bell inequalities in terms of actual properties of 
existing (partial) extensions rather than in terms of incompatibility 
between assumed principles. 
In particular, this will allow for a definite answer to the question 
\textit{what precisely is \lq\lq influenced\rq\rq } in arguments on
non-local effects in QM.


The extension problem outlined above has been studied in general in \cite{B-M}.
One of the result (see Sect.1) is that classical 
representability \textit{always holds} for yes/no observables 
with compatibility relations described by \textit{a tree graph}, i.e. a graph 
in which any two points are connected by exactly one path,
with points representing observables and links predicted correlations. 

In the case of four yes/no observables, $A_1,A_2,B_1, B_2$, 
$A_i$ compatible with $B_j$, this implies that, for fixed $B$, 
any pair of probabilities for the subsystems 
$A_i, B$ $i = 1,2$,  assigning the same probability to the 
outcomes of $B$, admits extensions to a probability 
on the whole system $A_1,A_2,B$, each giving a definite value to the 
possibly unmeasurable correlation $\langle A_1 A_2\rangle$. 
As we will see, the results on tree graphs also imply that 
\textit{if a value can be assigned to} $\langle A_1 A_2\rangle$ 
consistent with two choices $B_1$, $B_2$ for $B$,
then the whole system $A_i, B_j$ admits a classical representation.

Since the converse is obvious, it follows 
that the existence of a classical representation for given set  of 
expectations $\langle A_i\rangle $, $\langle B_i\rangle$ and  
correlations $\langle A_i B_j\rangle$ 
is equivalent to the possibility of giving a value to
$\langle A_1 A_2\rangle$ which is 
\textit{consistent with the two choices for $B$}.

The analysis of QM models with four yes/no observables is therefore reduced 
in general to the consistency (i.e. a non void intersection) 
between the ranges of the (hypotetical, unmeasurable) values 
for the correlation $\langle A_1 A_2\rangle$ allowed by (measurable) given 
correlations of $A_1$ and $A_2$ with $B_1$ and $B_2$.  

With respect to the same conclusion obtained by Fine \cite{Fine}, we 
stress that probabilistic descriptions of three-observable 
subsystem \textit{automatically exist} and only their compatibility 
is in question; moreover, only a repeated application of the results for 
tree graphs is required by our argument.
 

In Sect.2 , the possible range for $\langle A_1 A_2\rangle$, 
depending on the measurable correlations $\langle A_i B\rangle$, 
is characterized in terms of elementary three-observable 
inequalities. QM states which violate Bell inequalities 
are shown to give rise to disjoint intervals for the 
admissible values of $\langle A_1 A_2\rangle$ 
in automatically existing probabilistic models for $A_1,A_2,B_1$ 
and $A_1,A_2,B_2$.
This clearly shows that classical representability of
such states exactly fails in the attribution of a value to an 
unmeasurable correlation.

Actually, since BCH inequalities, more precisely, the  
eight inequalities discussed by Fine \cite{Fine}, 
are equivalent to classical representability, 
their violation exactly amounts to the inconsistency of such 
an attribution.

The same discussion and result apply to the states introduced, on the 
same set of observables, by Hardy \cite{Hardy} and  
exploited in Stapp's work \cite{Stapp1}, discussed by Mermin \cite{Mermin}.
In particular, Stapp's and Mermin's discussion exactly concerns 
partial extensions and the origin
of the inconsistency of the ranges of value for
an unmeasurable correlation $\langle A_1 A_2\rangle$ 
associated as above to the correlations of $A_i$ 
with $B_1$ and $B_2$ respectively. 


In Sect.3, the experiment proposed by Greenberger, Horne, 
Shimony and Zeilinger (GHSZ) \cite{GHSZ} is analyzed along the same lines.
The QM correlations defined by the GHSZ state, 
between observables $A_i$, $B_i$, $C_i$, $i = 1,2$,
are shown to extend to \textit{all correlations} within the sets of observables 
$A_i, B_i, C_1$ and  $A_i, B_i, C_2$; both extensions give 
\textit{unique values} for an \textit{ unmeasurable correlation}, 
$\langle A_1 A_2 B_1 B_2 \rangle = \pm 1$, and the sign depends  
on the choice between $C_1$ and $C_2$.  
Again, only the attribution of a value to an unmeasurable correlation
depends on the choice of an additional observable. 

In the last Section we will comment on the fact that such a dependence 
is the only basis for arguments about non local and non causal effects.

\section{\label{sec:ext}Extensions of QM predictions}

A basic notion is that of yes/no observables, denoting physical devices
producing, in each measurement, i.e. in each application to a physical system, 
two possible outcomes; by \textit{experimental setting} we will denote a 
collection of yes/no observables. 

Within an experimental setting, an \textit{(experimental) context} will 
denote a (finite) set of observables 
which \lq\lq  can be measured together 
on the same physical system\rq\rq, in the precise sense that
their joint application to a physical system 
gives rise to a statistics of their joint 
outcomes given by a classical probability,
i.e., by a normalized measure on the 
Boolean algebra freely generated by them. 
Observables appearing together in some context will be called 
\textit{compatible}; their correlations will be called 
\textit{measurable}, or \textit{observable}.

In the following, Quantum Mechanics will be interpreted as a theory 
predicting probabilities associated to experimental contexts defined
by sets of commuting projections. 
Joint statistics of non compatible observables are not defined in the above 
setting; correlations between incompatible observables will be called 
\textit{unmeasurable}, or \textit{unobservable}.

In particular, we will consider the 
\textit{Bell-Clauser-Horne (BCH) experimental setting}, 
consisting of two pairs ($A_1$, $A_2$), ($B_1$, $B_2$), of incompatible 
\textit{yes/no} observables, taking values in $\{ 0,1 \}$; 
the $A_i$ are compatible with the $B_j$, 
i.e. they can be measured together, in the sense introduced above. 
Such variables may be interpreted in terms of
spin components or photon polarizations, each pair referring to one
of a pair of particles, possibly in space-like separated regions. 

The experimental contexts given by QM consist, in the BCH setting, 
of the four pairs $\{A_i, B_j\}$; probabilities are associated to 
contexts, see Proposition \ref{prop:mis} below, by the observed relative 
frequencies $\langle A_i\rangle$, $\langle B_j\rangle$,
and $\langle A_i B_j\rangle$.
QM predictions for the BCH setting precisely consist 
in probabilities on contexts,
given by the spectral representation for the 
corresponding sets of commuting projectors. 

Such a \textit{partial} nature of QM predictions is a very general fact since
the (ordinary) interpretation of QM precisely consists of a set of probability 
measures, each defined, by the spectral theorem, 
on the spectrum of a commutative
subalgebra of operators in a Hilbert space.
Such a structure has been formalized in \cite{B-M} as a 
\textit{partial probability theory} on a \textit{partial Boolean algebra}. 

A basic problem for the interpretation of QM is \textit{the necessity} of such
partial structures; in other words, \textit{whether they admit a classical 
representation}, i.e. common probabilistic classical description in 
a probability space 
$(X,\Sigma, \mu)$, where $X$ is a set, $\Sigma$ a $\sigma$-algebra of 
measurable (with respect to $\mu$) subsets and $\mu$ is a probability 
measure on $X$ and all (yes/no) variables are described
by characteristic functions of measurable subsets.
Clearly, such a notion covers all kinds of \lq\lq hidden variables\rq\rq\ 
theories, all ending in the attribution of values to all the observables, with
definite probabilities. 

Extensions have been discussed in general in \cite{B-M}. Non trivial
extensions, describing QM predictions through a reduced number of contexts,  
have been shown to arise in some generality and some of the results 
can be described in terms of graphs: 

\begin{prop}\label{prop:tree1} 
Consider any set of probabilities $p_i$ on a set of yes/no
  observables $A_i$ and correlations $p_{ij}$ on a subset of pairs $A_i, A_j$,
defining a probability on each pair 
$A_i, A_j$, with $ p_i = \langle A_i \rangle$, 
$ p_j = \langle A_j \rangle$, $ p_{ij} = \langle A_i A_j \rangle$;
describing observables $A_i$ as points and the above pairs as links 
in a graph, any subset of predictions associated to \textit{a tree subgraph} 
admits a classical representation.
\end{prop}
 
\begin{prop}\label{prop:tree2} the same holds with 
\textit{yes/no observables} $A_i$ substituted by
free Boolean algebras $ \mathcal{A}_i $, 
$p_i$  by probabilities on 
$ \mathcal{A}_i $, $p_{ij} $ by probabilities on the Boolean algebra freely 
generated by the union of the sets of generators of 
$ \mathcal{A}_i $ and $ \mathcal{A}_j $. 
\end{prop}

Propositions \ref{prop:tree1} and \ref{prop:tree2} are proven in \cite{B-M}
by induction on the number of links of the correlation tree, by an explicit
construction of a probability on the Boolean algebra generated by two 
algebras in terms of conditional probabilities 
with respect to their intersection. The resulting probabilities are 
in general not unique.

The above Propositions can be interpreted as providing automatic
conditions for the realization of the possibility advocated by 
Einstein, Podolsky and Rosen \cite{EPR} to attribute a value to 
correlations between incompatible observables consistently 
with the QM predictions for the remaining, measurable, correlations; 
actually, they do not make reference to QM and give conditions
on compatibility relations allowing the extension of any set of 
predictions.

It is also useful to recall the following elementary fact \cite{B-M}, 
which implies that it is sufficient to analyze correlations since they 
completely define a probability measure.

\begin{prop}\label{prop:mis}
In a classical representation for $n$ observables $A_1,\ldots ,A_n$, the 
probability measure is completely defined by all expectation values 
$\langle A_i \rangle$ together with all possible correlations 
$\langle A_i A_j\rangle$,
$\langle A_i A_j A_k\rangle$,\ldots, $\langle A_1\ldots A_n\rangle$.
\end{prop}

In the BCH setting, the two subsets
$\{A_1, A_2, B_1\}$, $\{A_1, A_2, B_2\}$, with the correlations
$\langle A_j B_i \rangle$ $j=1,2,$ 
are described by tree graphs (see Fig.1 (a),(b))
and admit 
therefore classical representations. 
It follows that all QM predictions can be reproduced by a pair of
classical probability theories, one for 
the Boolean algebra freely generated by
$A_1, A_2, B_1$ the other for  $A_1, A_2, B_2$; 
both sets of variables include $A_1$ and $ A_2$ and give 
rise to predictions for the unmeasurable
correlation $\langle A_1 A_2 \rangle$.

In other terms, \textit{given a choice for} $B$, one 
may consistently speak of the correlation $\langle A_1 A_2 \rangle$.
Clearly, the existence of a classical representation for all the predictions
in the BCH setting implies the existence of a value for 
$\langle A_1 A_2 \rangle$ compatible with 
the different choices of $B$.

On the other hand, if it is possible to give a value to 
$\langle A_1 A_2 \rangle$ 
independently of the choice $B_1$, $B_2$ for $B$, this defines, 
by Proposition \ref{prop:mis}, a 
probability on the Boolean algebra $\mathcal{A}_{12}$ generated by $A_1, A_2$; 
then, the application of Proposition \ref{prop:tree2} 
to $\mathcal{A}_{12}$,  $\{B_1 \}$,  $\{B_2 \}$
implies classical representability of all the predictions for the 
BCH setting (see Fig.1 (c)). 
\begin{figure}
\begin{center}
\includegraphics[width=0.7\textwidth]{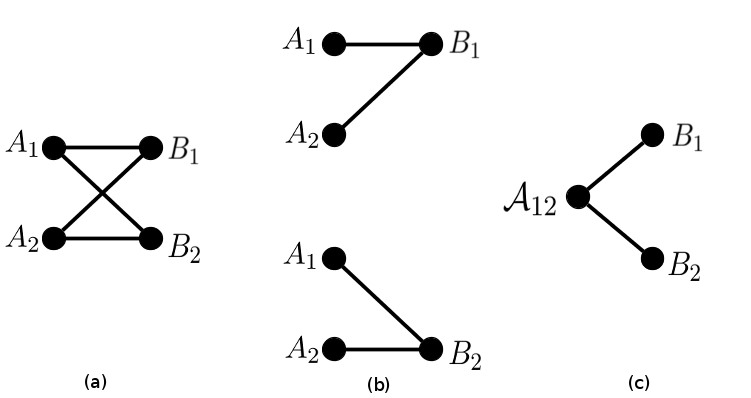}
\caption{(a) BCH compatibility graph; 
(b) two tree subgraphs; (c) tree graph corresponding to
extension of the measure on the algebra generated by $A_1,A_2$ }
\end{center} 
 \end{figure}

Classical representability and BCH inequalities 
(in the sense of Ref.\cite{Fine}) are therefore 
equivalent to the possibility of attributing a common value to the
unmeasurable correlation $\langle A_1 A_2 \rangle$ 
in classical models for $A_1, A_2, B_1$ and $A_1, A_2, B_2$.

In the next Section, we shall compute explicitly
the range of $\langle A_1 A_2 \rangle$ 
allowed by the QM predictions for a pair of spin variables, in the zero total
angular momentum state and for the state introduce by Hardy \cite{Hardy}.

\section{\label{sec:constr}Explicit Constraints 
for systems of three and four observables}

In this section  we first recall simple constraints, for three
 $ \{ 0, 1 \} $ valued random variables $ A_1, A_2, B $, 
satisfied by $\langle A_1 A_2\rangle$ given  
$\langle A_i \rangle$, $\langle B \rangle$ and 
$\langle A_i B \rangle$. We then show that, for the 
predictions given by both the QM states discussed 
by Bell and Hardy \cite{Hardy}, such constraints give 
rise, for different choices of $B$, to incompatible 
values for $\langle A_1 A_2\rangle$. 

The possible ranges of correlations in a classical model
for $\{A_1,A_2,B\}$ are given by the Bell-Wigner 
polytope \cite{Pit89}, also discussed in \cite{B-M}. 
We only need a subset of the corresponding set of 
inequalities, namely
\begin{equation}
\label{P1} 
0 \leq \langle A_1 A_2\rangle \leq \langle A_1 \rangle
\end{equation}
\begin{equation}
\label{P2} 
\langle A_1 B\rangle + \langle A_2B\rangle -\langle B \rangle 
\leq \langle A_1 A_2 \rangle  
\leq \langle A_2 \rangle - \langle A_2B \rangle +\langle A_1B \rangle   
\end{equation}
\begin{equation}
\label{P3} 
\langle A_1\rangle +\langle A_2\rangle +\langle B\rangle -
\langle A_1B \rangle -\langle A_2 B\rangle - 1  
\leq \langle A_1 A_2\rangle   
\end{equation}
Eq. (\ref{P1}) is trivial; the first relation in eq. (\ref{P2}) 
immediately follows from 
$ \langle (A_1 - B)(A_2 - B) \rangle \geq 0$, which holds 
since for $B=1$ both factors are non positive, for $B=0$ 
they are both non negative; moreover, the first relation 
in eq. (\ref{P2}) implies the second by interchanging $B$ and $A_2$ 
and eq. (\ref{P3}) by interchanging $B$ and $1-B$.
The complete set of constraints \cite{Pit89} includes three
additional inequalities; they are obtained from the above relations 
through the interchange of $A_1$ and $A_2$ 
and the substitution of $B$ with $1$ in eqs. (\ref{P1}), (\ref{P2})
(and are not relevant for our analysis).

The constraints given by eqs. (\ref{P1}), (\ref{P2}) (\ref{P3}), 
applied to QM predictions for mean values and \textit{observable} 
correlations, may give rise to  ranges of values for the 
unobservable correlation $ \langle A_1 A_2 \rangle $ which are disjoint
for different choices of $B$. 

Consider in fact a system of two spin $1/2$ particles in the singlet state
${\frac{1}{\sqrt{2}}(\lvert+-\rangle- \lvert-+\rangle)}$ 
and associate $A_1$ and $A_2$ with the
projectors on the spin up state along directions $\hat{x}$ and
$\hat{z}$ for the first particle and $B_1$ with the projector on 
the spin down state along the direction
$\frac{\hat{x}+\hat{z}}{\sqrt{2}}$ for the second particle, namely
\begin{equation}
\frac{1+\sigma_x}{2},\qquad \frac{1+\sigma_z}{2},
\qquad \frac{1}{2}-\frac{\tau_x+\tau_z}{2\sqrt{2}},
 \end{equation}
where $\sigma_i$'s and $\tau_i$'s are Pauli matrices acting,
respectively, on the first and on the second subsystem.
Quantum mechanical predictions give
\begin{align}\label{eq:val1}
 \langle A_1\rangle = \langle A_2\rangle = \langle  B_1\rangle = 
\frac{1}{2}  \ ,\\ 
\label{eq:val13} \langle A_1 B_1 \rangle = 
\langle A_2 B_1\rangle = \frac{1}{4} +\frac{\sqrt{2}}{8} \ , 
 \end{align}
Denoting by  $\langle\text{ }\rangle_1$ correlations in
a probabilistic model for $\{A_1,A_2,B_1\}$ eqs. (\ref{P1}) and (\ref{P2}) 
give 
\begin{equation}\label{eq:rng1}
  \frac{1}{4}  < \frac{\sqrt{2}}{4}\leq
\langle A_1 A_2\rangle_1\leq \frac{1}{2} \ .
\end{equation}
If $B_1$ is substituted by $B_2$, representing the projector
\begin{equation}
\frac{1}{2}-\frac{\tau_x-\tau_z}{2\sqrt{2}} \ ,
\end{equation}
QM predictions give $\langle B_2 \rangle=\frac{1}{2}$ and
\begin{equation}
\label{eq:val13'} \langle A_1 B_2\rangle =\frac{1}{4}+\frac{\sqrt{2}}{8} \  , 
 \quad \langle A_2 B_2\rangle = \frac{1}{4}-\frac{\sqrt{2}}{8} \ .
\end{equation}
Eqs. (\ref{P1}) and (\ref{P2}) then give, for the unmeasurable
correlation $\langle\text{ }\rangle_2$ in
probabilistic models for $\{A_1,A_2,B_2\}$ 
\begin{equation}\label{eq:rng2}
 0\leq \langle
 A_1 A_2 \rangle_2\leq\frac{1}{2}-\frac{\sqrt{2}}{4} < \frac{1}{4} \ .
\end{equation}
It follows that the value of $\langle
A_1 A_2\rangle$ must belong to one of two 
\textit{disjoint intervals}, depending
on whether one is considering the probability space for $\{A_1,A_2,B_1\}$
or that for $\{A_1,A_2,B_2\}$, i.e. whether one chooses to measure $B_1$ or
$B_2$ on the second subsystem. 
Therefore, the validity of QM predictions for compatible observables 
\textit{does not allow} to assign a definite value to the
correlation $\langle A_1 A_2\rangle$, independent of the choice of the 
measurement to be performed on the  second subsystem. 

The second example is Hardy's experiment \cite{Hardy}, 
also exploited in Stapp's \cite{Stapp1} and Mermin's \cite{Mermin} 
discussion; we shall refer to Mermin's notation.
As in the previous example, we have two spin $1/2$ particles and four yes/no 
observables, denoted by $L1,L2$ and $R1,R2$, acting respectively on a 
``left'' and a ``right'' particle. 
The state is described, up to a normalization 
factor, by the vector
\begin{equation}
\lvert \Psi\rangle = \lvert L1+,R1-\rangle - 
\lvert L2-, R2+\rangle \langle L2-, R2+\lvert L1+, R1-\rangle\ ,
\end{equation}
where, e.g,  $\lvert L1+,R1-\rangle$ 
indicates a simultaneous eigenstate of the commuting
observables $L1$ and $R1$ with eigenvalue $1$ on the left and $-1$ on the right.

The variables $A_i$, $i=1,2$, are associated with the propositions 
``the result of the measurement of $Ri$ is $+1$'', the same
for $B_i$ and $Li$; as before, $A_i$ and $B_i$ take values in $\{0,1\}$.
Independently of the specific expressions for $Li$ and $Ri$, 
QM predicts the following correlations   
(corresponding to eqs.(6)-(9) in Ref. \cite{Mermin}):
\begin{eqnarray}
\label{eq:s1}  1-\langle A_2 \rangle -\langle B_1 
\rangle+\langle A_2 B_1 \rangle =    0 \ ,\\
\label{eq:s2}   \langle A_2 \rangle -\langle A_2 B_2 \rangle = 0 \ ,\\
\label{eq:s3}   \langle A_1 B_2\rangle = 0 \ ,\\
\label{eq:s4}   \langle A_1 \rangle - \langle A_1 B_1 \rangle > 0 \ ,
\end{eqnarray}
In eq. (\ref{eq:s4}), $>0$ 
is equivalent to $\neq 0$ in Ref. \cite{Mermin} since 
$\langle A_1\rangle\geq \langle A_1 B_1 \rangle$ always holds.

Eqs. (\ref{P3}), (\ref{eq:s1}) and (\ref{eq:s4}) give, 
in all classical models for $\{A_1,A_2,B_1\}$ 
\begin{equation}
\langle A_1 A_2\rangle_1 \geq \langle A_1\rangle +
\langle A_2\rangle +\langle B_1\rangle -\langle A_1B_1
\rangle -\langle A_2 B_1\rangle -1 = \langle A_1\rangle -\langle A_1B_1
\rangle > 0 \ ;
\end{equation}
on the other hand, in classical models for $\{A_1,A_2,B_2\}$, 
eqs. (\ref{P2}), (\ref{eq:s1}) and (\ref{eq:s4}) give, 
\begin{equation}
\langle A_1 A_2\rangle_2 \leq \langle A_2\rangle -\langle A_2B_2\rangle  + 
\langle A_1B_2 \rangle = 0 \ ,
\end{equation}
which also implies, by eq. (\ref{P1}), 
\begin{equation}
\langle A_1 A_2\rangle_2 = 0 \ .
\end{equation}

As in the previous example, no value for the unmeasurable 
correlation $\langle A_1 A_2\rangle$ is compatible 
with different choices of $B_i$. 
A peculiarity of this case is that in classical models for 
$\{A_1,A_2,B_2\}$ the correlation $\langle A_1 A_2\rangle$ is completely fixed 
(to $0$) by the constraints imposed by measurable correlations.
Attributing the $0$ value is actually equivalent to asserting the logical 
implication $A_1 \rightarrow A_2^c$, i.e. 
``\textit{if} the measurement of $R1$ gives $+1$, 
\textit{then} the measurement of $R2$ gives $-1$''; such an implication   
\textit{holds in all classical models for} 
$\{A_1,A_2,B_2\}$ \textit{and fails   
in all classical models for} $\{A_1,A_2,B_1\}$.

\section{\label{sec:GHSZ}The perfect correlation model of 
GHSZ}

In this Section, the above analysis is applied to the case considered 
in \cite{GHSZ} and discussed in \cite{Mermin2} within a general 
framework unifying Bell and Kochen-Specker kinds of results.

With a slight modification of the above terminology, 
the experimental setting involved in the GHSZ model consists 
of six yes/no observables, which can be described as $A_i$, $B_i$, $C_i$,
$i = 1,2$, 
taking values $1$ and $-1$; equivalently, by the propositions
$P_i$, $Q_i$, $R_i$ respectively asserting 
$A_i = 1$, $B_i = 1$, $C_i = 1$.
The experimental contexts, i.e. the sets of compatible observables 
are given by the possible choices of (at most) one $A$, one $B$ and one $C$. 

In Refs. \cite{GHSZ} and \cite{Mermin2} observables associated to 
different letters are interpreted as measured in distant, 
space-like separated regions.
The above observables reproduce the example of Ref. \cite{GHSZ}, Sect. III, 
with the identification $ A_i = A(\phi_i)B(0)$, $B_i = C(\phi_i)$, 
$C_i =D(\phi_i)$, with $\phi_1 = 0$ $\phi_2= \pi/2$; 
in Mermin's spin notation, they should be read as
$A_1 = - \sigma^1_x$,   $A_2 = - \sigma^1_y$,   
$B_1 =  \sigma^2_y$,   $B_2 =  \sigma^2_x$,   
$C_1 =  \sigma^3_y$,   $C_2 =  \sigma^3_x$.

In general, quantum mechanical predictions consist in probability 
assignments within each of the eight contexts defined by the choice of
three indexes $i$; e.g. a context is given by the choice
$A_1$, $B_1$, $C_1$, and the associated predictions by a probability on the
Boolean algebra freely generated by the propositions 
$P_1$, $Q_1$, $R_1$.

The crucial point of GHSZ and Mermin's analysis is the observation that, for 
suitable states, QM predictions give \lq\lq perfect correlations\rq\rq\
(each correlation involving observables in a fixed context), 
which are not compatible with any
(context-independent) assignment of values to the variables 
$A_i$, $B_i$, $C_i$. 

In fact, the state considered by GHSZ and Mermin can be written, for the
Mermin spin variables $\sigma^i_k$, 
in the usual notation referring to
the eigenvalues of $\sigma^i_3$, 
\begin{equation}
\label{psi}
\lvert \Psi \rangle = \frac{1}{\sqrt 2} \ (\lvert +++\rangle - \lvert---\rangle)
\end{equation}
and gives the correlations
\begin{eqnarray}
\label{perfcorr}
\langle A_1 B_1 C_1 \rangle = -1\ , \qquad \langle A_2 B_2 C_1\rangle
= -1\ , \\ 
\langle A_2 B_1 C_2 \rangle = -1\ , \qquad 
\langle A_1 B_2C_2\rangle = +1\ . \label{perfcorr2}
\end{eqnarray}
Values $a_i, b_i, c_i = \pm 1$ assigned to the above variables and 
reproducing the above correlations would satisfy the relations  
\begin{equation}
a_1 b_1 c_1 = -1\ , \ \ a_2 b_2 c_1= -1\ , \ \ a_2 b_1 c_2 = -1\ , 
\ \ a_1 b_2 c_2 = +1\ ,
\end{equation}
which have no solution since their product gives $1$ for the l.h.s.,
$-1$ for the r.h.s..

Following the program outlined above and using the notions 
and results of Sect. 1, 
we will extend the GHSZ and Mermin argument to show: 
\begin{itemize}

\item[$i)$] the correlations eqs. (\ref{perfcorr}), (\ref{perfcorr2}) 
\textit{and all the other QM
predictions} given by the GHSZ state, eq. (\ref{psi}), 
extend to classical probabilities $p_1$, $p_2$ 
on the \textit{two} subsystems defined by  
$\{A_1, A_2, B_1, B_2, C_1\}$ and $\{A_1, A_2, B_1, B_2, C_2\}$ 

\item[$ii)$] unique and different values, $\pm1$, are given by such 
$p_1$ and $p_2$ to the correlation $\langle A_1 A_2 B_1 B_2 \rangle $, 
all the other correlations between the $A$ and $B$ observables
admitting common values;

\item[$iii)$] pairs of probabilities on the two subsystem 
introduced in $i)$ extend to a probability on 
the entire system generated by      
$A_i, B_j, C_k$, $i,j,k, = 1,2$, 
exactly if they coincide on the subsystem generated by 
$A_1,A_2,B_1,B_2$.
\end{itemize}

Again, the conclusion is that \textit{exactly a non observable
correlation}, $\langle A_1 A_2 B_1 B_2 \rangle $, \textit{depends} on the 
choice of the observable $C_i$.

In order to derive $i)$ - $iii)$, 
we first observe that $iii)$ follows immediately
from Proposition \ref{prop:tree2}, applied to the Boolean algebra generated by  
$P_i, Q_j$ and to its correlations with the observables 
$R_1$ and $R_2$, forming a tree graph.

Concerning $ii)$, observe that  eqs. (\ref{perfcorr}) imply,
for any probabilistic model reproducing them, 
$A_1 B_1 C_1 = -1 $ and $ A_2 B_2 C_1 = -1 $ with probability $1$,
 so that, with probability $1$, 
$ A_1 B_1  A_2 B_2  = 1 $; equivalently,
$ \langle A_1 B_1 A_2 B_2 \rangle = 1 $.
In the same way,  eqs. (\ref{perfcorr2}) imply
$ \langle A_1 B_1 A_2 B_2 \rangle = - 1 $; such relations hold therefore
for all probabilistic models reproducing, respectively, the QM correlations 
of the $A,B$ observables with $C_1$ and $C_2$. The existence of probabilities
giving common values to all the other correlations follows from the
construction below.

The most involved issue is $i)$, which is non trivial since quantum mechanical
predictions involve eight different contexts and $i)$ states that the
predictions given by the state (\ref{psi}) extend 
to \textit{all correlations} between variables inside each one of 
\textit{only two} contexts.
Notice that Propositions \ref{prop:tree1} and \ref{prop:tree2} 
do not imply such an extension, since 
the corresponding set of correlations is not given by a tree graph and in fact
not all quantum states admit it, as also implied by the 
BCH analysis; the exact form of the quantum predictions we are going 
to extend is therefore important. It is easy to see that the state given 
by eq. (\ref{psi}) gives rise to the correlations
\begin{equation}
\label{otherpred1}
\langle A_i \rangle = \langle B_i \rangle = \langle C_i \rangle =
\langle A_i B_j \rangle = \langle A_i C_j\rangle = \langle B_i C _j
\rangle =0 \ , \ i,j=1,2
\end{equation}
\begin{equation}
\label{otherpred2}
 \langle A_1 B_2 C_1 \rangle = \langle A_2
B_1 C_1 \rangle = \langle A_2 B_2 C_2 \rangle= \langle A_1 B_1 C_2
\rangle = 0 \ .
\end{equation}
We have to show that the all the correlations given by eqs.
(\ref{perfcorr}),(\ref{perfcorr2}),(\ref{otherpred1}),(\ref{otherpred2}) 
involving
the observables $A_i, B_j, C_1$ 
can be represented by a classical model,
and the same for those involving $A_i, B_j, C_2$, and that the two models
give the same correlations between variables
$A_i, B_j$ with the only exception of
$\langle A_1 A_2 B_1 B_2 \rangle $.

The first classical model is defined as follows: let $A_i, C_1$ 
denote variables
taking values $1$ and $-1$ with probability $1/2$, i.e.
\begin{equation}
\label{mod11} 
A_i^2 = 1\, ,\  C_1^2 = 1 \ , \ \
   \langle A_i \rangle = \langle  C_1 \rangle = 0 \ , \ \ 
   \langle A_1  A_2 \rangle = \langle A_i  C_1 \rangle = 0 
\end{equation}
and define 
\begin{equation}
\label{mod12} 
 B_i \equiv - A_i C_1 \ .
\end{equation}
Eqs. (\ref{mod11}) and (\ref{mod12}) immediately imply eqs. (\ref{perfcorr}) 
and all eqs. (\ref{otherpred1}),(\ref{otherpred2}) 
not involving $C_2$.
 
The second model is defined in the same way, by independent variables
$A_i$ and $C_2$ satisfying the same relations as in eqs. (\ref{mod11}) 
and by variables $B_i$ now defined as 
\begin{equation}
\label{mod22} 
 B_1 \equiv - A_2 C_2 \  , \ \  B_2 \equiv  A_1 C_2 \ .   
\end{equation}
Eqs. (\ref{perfcorr2}) and all eqs. (\ref{otherpred1}), (\ref{otherpred2}) 
not involving $C_1$ immediately follow.

All the correlations defined by the two models between $A$ and $B$ variables 
follow from eqs. (\ref{mod11}), with $C_2$ replacing $C_1$ for the second model, 
(\ref{mod12}),(\ref{mod22}); in both models
$\langle A_1  A_2 \rangle =  0 $ by definition,
\begin{equation}
\label{modsother}    
\langle B_1  B_2 \rangle = \pm \langle A_1  A_2  \rangle = 0 
\end{equation}
and
$$
\langle A_i B_1  B_2 \rangle = 0 =  \langle B_i A_1  A_2  \rangle  \ ,
$$
both expectations reducing to $ \pm \langle A_k \rangle $ for some $k$.
The two models give therefore identical predictions for all the correlations
between the $A$ and $B$ variables, with the only exception 
\begin{equation}
\label{modsdiff}    
\langle A_1 A_2 B_1 B_2 \rangle = \pm 1 \ .
\end{equation}

\section{\label{conclusions}Conclusions}

The discussion of Bell inequalities usually begins with their
derivation from general \textit{locality} and \textit{reality} 
principles. Reality principles \textit{include} hypothetical 
common attributions of values to observables even when they are not 
compatible according to QM; once such attributions are assumed, 
any logical or probabilistic consideration automatically concerns
extensions of QM predictions to unmeasurable correlations between 
incompatible observables.

Moreover, arguments about the possibility of considering
measurements which \textit{could have been performed}
and the assumption that the results of such \lq\lq unperformed\rq\rq\
measurements satisfy the predictions of QM for their correlation with
other, compatible and performed, measurements, lead  
\cite{Stapp1} \cite{Mermin}
to logical or probabilistic relations 
between quantum mechanically incompatible observables
which precisely amount 
to partial extensions of QM predictions.

We have shown that, independently of any additional principle 
or argument, partial extensions of QM predictions are 
always possible in cases described by tree graphs,
that the ranges of values for correlations between 
quantum mechanically incompatible observables obtained by 
such extensions depend in general on the considered set of observables 
and that they may be incompatible for different sets.

As shown by the Hardy and the GHSZ examples, this may happen 
even when \textit{all extensions}, 
to a fixed allowed set of observables, 
of a set of QM predictions give the same 
\textit{perfect correlations} to quantum mechanically incompatible 
observables.
If the \textit{dependence of such correlations on a set
of observables} allowing for a partial extension 
is not taken into account, the result appears as a 
contradiction between logical consequences of QM predictions 
(and of experimental results).

Moreover, precisely unmeasurable correlations depend on
a choice in a far or future space-time regions in the discussion 
of locality and causality principles.
In fact, in the BCH setting, identifying $B_i$ with observables measured 
in a space-like separated region, only and exactly 
$\langle A_1 A_2 \rangle $ depends on the choice of $B_i$; 
alternatively, if $B_i$ is measured in a future region,
such a dependence violates causality
in the precise sense that any hypotetical record of 
the value of $\langle A_1 A_2\rangle$ depends on a future choice. 

All the above \lq\lq violations\rq\rq\ exactly concern  
extensions of QM predictions to \textit{correlations 
between quantum mechanically incompatible observables}, e.g., between 
two different spin components of \textit{the same} particle or polarization 
directions of \textit{the same} photon. 

A discussion of the violation of BCH inequalities in terms of 
unobservable correlations  has also been given in Ref. \cite{Cohen}. 
A (different, but related) notion of \lq\lq extension of 
quantum correlations\rq\rq\ is also central 
in the recent discussion of the Einstein-Podolsky-Rosen notion of 
reality in Ref. \cite{Nistico}.

\end{document}